\documentstyle[11pt,epsfig]{article}
\title{\bf Perihelion advance and trajectory of charged test particles in Reissner-Nordstrom field via the higher-order geodesic deviations}
\author{{M. Heydari-Fard \thanks{Electronic address: heydarifard@qom.ac.ir}, S. Fakhry \thanks{Electronic address: s\_fakhry@sbu.ac.ir} and S. N. Hasani \thanks{Electronic address: n.hasani@stu.qom.ac.ir }}\\ {\small \emph{Department of Physics, The University of Qom, P.
O. Box 37185-359, Qom, Iran}}
\\{\small \emph{Department of Physics, Shahid Beheshti University, G. C., Evin, Tehran 19839, Iran}}}
\begin{document}
\maketitle %\baselineskip 24pt
\begin{abstract}
By using the higher-order geodesic deviation equations for charged particles, we apply the method described by Kerner et.al. to calculate the  perihelion advance and trajectory of charged test particles in the Reissner-Nordstrom space-time. The effect of charge on the perihelion advance is studied and compared the results with those obtained earlier via the perturbation method. The advantage of this approximation method is to provide a way to calculate the perihelion advance and orbit of planets in the vicinity of massive and compact objects without considering Newtonian and post-Newtonian approximations.
\vspace{5mm}
\vspace{1mm}
\end{abstract}

\section{Introduction}
The problem of planets motion in general relativity is the subject of many studies in which the planet has been considered as a test particle moving along its geodesic \cite{Einstein}. Einstein made the first calculations in this regard for the planet Mercury in the Schwarzschild space-time which resulted in the equation for the perihelion advance
\begin{equation}
\Delta{\varphi}=\frac{6\pi GM}{a(1-e^{2})},\label{00}
\end{equation}
where $G$ is the gravitational constant, $M$ is the mass of the central body, $a$ is the length of semi-major axis for planet's orbit and $e$ is eccentricity. Derivation of perihelion advance by using this method leads to a quasi-elliptic integral whose calculation is very difficult, which is then evaluated after expanding the integrand in a power series of the small parameter $GM/rc^2$. For the low-eccentricity trajectories of planets, one can obtain the following approximate formula for the perihelion advance
\begin{equation}
\Delta{\varphi}=\frac{6\pi GM}{a(1-e^{2})}\simeq \frac{6\pi GM}{a} (1+e^{2}+e^{4}+e^{6}+\cdots),\label{1}
\end{equation}
even for the case of Mercury up to second-order of eccentricity, the perihelion advance differs only by $0.18\%$ error from its actual value \cite{Kerner}. It should be noted again that Einstein's method is only valid for the small values of $GM/rc^2$.

In what follows, we show that one can obtain the same results (without taking the complex integrals) only by considering the successive approximations
around a circular orbit in the equatorial plane as the initial geodesic with constant angular velocity, which leads to an iterative process of the solving the geodesic deviation equations of first, second and higher-orders \cite{Kerner2, Colistete, Holten}. Here, instead of the $GM/rc^2$ parameter the eccentricity, $e$, plays the role of the small parameter which is controlling the maximal deviation from the initial circular orbit. In this method, we have no constraint on $GM/rc^{2}$ anymore. So, one can determine the value of perihelion advance for large mass objects and write it in the higher-order of $GM/rc^{2}$.

The orbital motion of neutral test particles via the higher-order geodesic deviation equations for Schwarzschild and Kerr metrics are studied in \cite{Kerner} and \cite{Colistete}, respectively. Also, for massive charged particles in Reissner-Nordstrom metric, geodesic deviations have been extracted up to first-order \cite{Balakin}. In this paper, by using the higher-order geodesic deviations for charged particles \cite{Heydari-Fard}, we are going to obtain the orbital motion and trajectory of charged particles. We also expect that our calculations reduce to similar one in Schwarzschild metric \cite{Kerner} by elimination of charge. In fact, we generalize the novel method used in reference \cite{Kerner} for neutral particles in the Schwarzschild metric to the charged particles in the Reissner-Nordstrom metric. Recently, an analytical computation of the perihelion advance in general relativity via the Homotopy perturbation method has been proposed in \cite{0}. Also, one can study the perihelion advance of planets in general relativity and modified theories of gravity by using different methods in \cite{1}--\cite{15}.

The structure of the paper is as follows: In section 2, by using the approximation method introduced in reference \cite{Heydari-Fard}, we derive the higher-order geodesic deviation for charged particles. By using the first-order geodesic deviation equations, the orbital motion of charged particles is found in section 3. In section 4, we obtain the second-order geodesic deviations and derive the semi-major axis, eccentricity, and trajectory using the Taylor expansion around a central geodesic. The obtained results are discussed in section 5.

\section{The higher-order geodesic deviation method}
\begin{figure}
\centerline{\begin{tabular}{ccc} \epsfig{figure=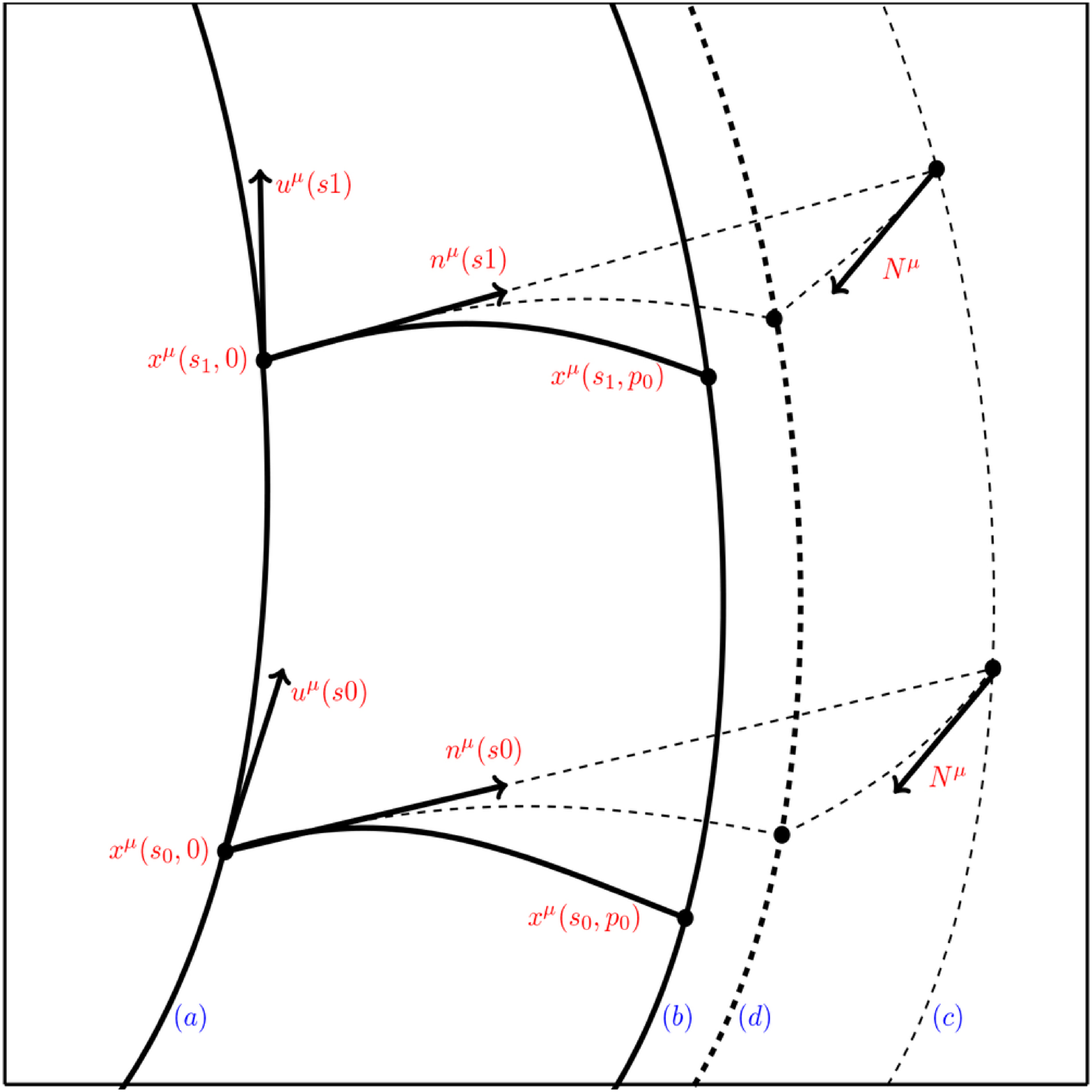,width=7cm}
\end{tabular} } \caption{\footnotesize  Deviation of two nearby geodesics in a gravitational field. Lines (a)
and (b) represent the central geodesic $p = 0$ and the nearby geodesic
$p = p_0$, respectively; lines (c) and (d) show the corresponding first
and second-order approximations to the nearby geodesic (b). Also, $u^{\mu}$ is the unit tangent vector to the central world-line, $n^{\mu}$ is the separation vector to the curve $s = const$ and $N^{\mu}=b^{\mu}-\Gamma^{\mu}_{\lambda \rho}n{^\lambda}n^{\rho}$ is the second-order geodesic deviation \cite{Baskaran}.}\label{fig1}
\end{figure}
As it mentioned above, the higher-order geodesic deviation equations for charged particles have been derived in \cite{Heydari-Fard} for the first-time. In this section, we are going to derive the geometrical set-up used in our work. The geodesic deviation equation for charged particles is \cite{Balakin}
\begin{equation}
\frac{D^{2}n^{\mu}}{Ds^{2}}=R^{\mu}_{\,\,\,\,
\lambda\nu\kappa}u^{\lambda}u^{\nu}n^{\kappa}+\frac{q}{m}F^{\mu}_{\,\,\,\,\nu}\frac{Dn^{\nu}}{Ds}
+\frac{q}{m}\nabla_{\lambda}F^{\mu}_{\,\,\,\,\nu}u^{\nu}n^{\lambda},\label{2}
\end{equation}
where $\frac{D}{Ds}$ is the covariant derivative along the curve and $n^{\mu}$ is the separation vector between two particular neighboring geodesics (see figure 1). Here, $u^{\mu}$ is the tangent vector to the geodesic, $R^{\mu}_{\,\,\,\,\lambda\nu\kappa}$ is the curvature tensor of space-time, $q$ and $m$ are charge and mass of particles (particles have the same charge-to-mass ratio, $q/m$), and $F^{\mu}_{\,\,\,\,\nu}$ is the electromagnetic force acting on the charged particles. For neutral particles, the above equation reduces to the following geodesic deviation \cite{Misner, Weinberg}
\begin{equation}
\frac{D^{2}n^{\mu}}{Ds^{2}}=R^{\mu}_{\,\,\,\,
\lambda\nu\kappa}u^{\lambda}u^{\nu}n^{\kappa},\label{0}
\end{equation}
which is the well-known equation (Jacobi equation) in general relativity. We introduce the four-velocity $u^{\alpha}(s, p)=\frac{\partial x^{\alpha}}{\partial s}$ as the time-like tangent vector to the world-line and $n^{\alpha}(s, p)=\frac{\partial x^{\alpha}}{\partial p}$ as the deviation four-vector as well. Practically it is often convenient to work with the non-trivial covariant form. It can be obtained by replacement of the trivial expressions for the covariant derivatives, the Riemann curvature tensor and use of the equation of motion in the left-hand side of equation (\ref{2}) \cite{Balakin}
\begin{equation}
\frac{d^{2}n^{\mu}}{ds^{2}}+\left(2\Gamma ^{\mu}_{\,\,\,\,\kappa\nu}u^{\kappa}-\frac{q}{m}F^{\mu}_{\,\,\,\,\nu}\right)\frac{dn^{\nu}}{ds}+\left(u^{\kappa}u^{\sigma}\partial _{\nu}\Gamma ^{\mu}_{\,\,\,\,\kappa\sigma}-\frac{q}{m}u^{\kappa}\partial _{\nu}F^{\mu}_{\,\,\,\,\kappa}\right)n^{\nu}=0.\label{3}
\end{equation}
The geodesic deviation can be used to compose geodesics $x^{\mu}(s)$ near a given reference geodesic $x^{\mu}_{0}(s)$, by an iterative method as follows. Considering this, one can write Taylor expansion of $x^{\mu}(s, p)$ around the central geodesic and obtain the first-order and higher-order geodesic deviations for charged particles
\begin{equation}
x^{\mu}(s, p)=x^{\mu}(s, p_{0})+(p-p_{0})\frac{\partial x^{\mu}}{\partial p}|_{(s, p_{0})}+\frac{1}{2!}(p-p_{0})^{2}\frac{\partial ^{2} x^{\mu}}{\partial p^{2}}|_{(s, p_{0})}+\cdots,\label{4}
\end{equation}
our aim is to obtain an expression in terms of the deviation vector. As shown in the above equation, the second term, $\frac{\partial x^{\mu}}{\partial p}$, is the definition of deviation vector and shows the first-order  geodesic deviation. But in the third term, $\frac{\partial ^{2} x^{\mu}}{\partial p^{2}}$, is not vector anymore. Therefore, we define the vector $b^{\mu}$ as follow
\begin{equation}
b^{\mu}=\frac{Dn^{\mu}}{Dp}=\frac{\partial n^{\mu}}{\partial p}+\Gamma ^{\mu}_{\lambda\nu}n^{\lambda}n^{\nu},\label{5}
\end{equation}
to change $\frac{\partial ^{2} x^{\mu}}{\partial p^{2}}$ into the expression showing the second-order geodesic deviation. By substituting equation (\ref{5}) into equation (\ref{4}), one can obtain the expression in terms of the order of vector deviation
\begin{equation}
x^{\mu}(s, p)=x^{\mu}(s, p_{0})+(p-p_{0})n^{\mu}+\frac{1}{2!}(p-p_{0})^{2}(b^{\mu}-\Gamma ^{\mu}_{\lambda\nu}n^{\lambda}n^{\nu})+\cdots\label{6}
\end{equation}
In the above expression, one can make some changes for simplification. We consider $\delta ^{n} x^{\mu}(s)$ as $n$-th order of geodesic deviation and by assuming $(p-p_{0})$ as a small quantity, $\epsilon$, we rewrite equation (\ref{6}) as follows
\begin{equation}
x^{\mu}(s, p)=x^{\mu}_{0}(s)+\delta x^{\mu}(s)+\frac{1}{2}\delta ^{2}x^{\mu}(s)+\cdots,\label{7}
\end{equation}
where $\delta x^{\mu}(s)=\epsilon n^{\mu}(s)$ is the first-order geodesic deviation and $\delta ^{2}x^{\mu}(s)=\epsilon^{2}(b^{\mu}-\Gamma^{\mu}_{\nu\lambda}n^{\nu}n^{\lambda})$ is the second-order geodesic deviation. In order to obtain the second-order geodesic deviation equation, one can apply the definition of the covariant derivative on equation (\ref{5})(for more details see reference \cite{Heydari-Fard} and appendix therein)
 \begin{eqnarray}
\frac{D^{2}b^{\mu}}{Ds^{2}}+R^{\mu}_{\,\,\rho\lambda\sigma}u^{\rho}u^{\lambda}n^{\sigma}&=&(R^{\mu}_{\,\,\rho\lambda\sigma ;\nu}-R^{\mu}_{\,\,\sigma\nu\rho ;\lambda})u^{\lambda}u^{\sigma}n^{\rho}n^{\nu}+4R^{\mu}_{\,\,\rho\lambda\sigma}u^{\lambda}\frac{Dn^{\rho}}{Ds}n^{\sigma}\nonumber\\
&+&\frac{q}{m}R^{\mu}_{\,\,\sigma\rho\nu}F^{\rho}_{\,\,\lambda}n^{\sigma}u^{\lambda}n^{\nu}+\frac{q}{m}F^{\mu}_{\,\,\rho ;\lambda\sigma}n^{\lambda}u^{\rho}n^{\sigma}+\frac{q}{m}F^{\mu}_{\,\,\rho}\frac{D^{2}u^{\rho}}{Dp^{2}}.\label{8}
\end{eqnarray}
Similar to the first-order geodesic deviation equation (\ref{3}), we can write equation (\ref{8}) in the non-manifest covariant form
\begin{eqnarray}
\frac{d^{2}b^{\mu}}{ds^{2}}&+&\left(2\Gamma^{\mu}_{\kappa\nu}u^{\kappa}-\frac{q}{m}F^{\mu}_{\,\,\,\,\nu}\right)\frac{db^{\nu}}{ds}
+\left(u^{\kappa}u^{\sigma}\partial_{\nu}\Gamma^{\mu}_{\kappa\sigma}-\frac{q}{m}u^{\kappa}\partial _{\nu}F^{\mu}_{\,\,\,\,\kappa}\right)b^{\nu}=\nonumber\\
&+&\left(\Gamma^{\tau}_{\sigma\nu}\partial_{\tau}\Gamma^{\mu}_{\lambda\rho}+2\Gamma^{\mu}_{\lambda\tau}\partial_{\rho}\Gamma^{\tau}_{\sigma\nu}-
\partial_{\nu}\partial_{\sigma}\Gamma^{\mu}_{\lambda\rho}\right)\left(u^{\lambda}u^{\rho}n^{\sigma}n^{\nu}-u^{\sigma}u^{\nu}n^{\lambda}n^{\rho}\right)\nonumber\\
&+&4\left(\partial_{\lambda}\Gamma^{\mu}_{\sigma\rho}+\Gamma^{\nu}_{\sigma\rho}\Gamma^{\mu}_{\lambda\nu}\right)\frac{dn^{\sigma}}{ds}(u^{\lambda}n^{\rho}
-u^{\rho}n^{\lambda})
+\frac{q}{m}u^{\nu}n^{\alpha}n^{\beta}\left(\partial_{\alpha}\partial_{\beta}F^{\mu}_{\,\,\,\,\nu}-
F^{\mu}_{\,\,\,\,\rho}\partial_{\nu}\Gamma^{\rho}_{\alpha\beta}-\partial_{\sigma}F^{\mu}_{\,\,\,\,\nu}\Gamma^{\sigma}_{\alpha\beta}\right)\nonumber\\
&+&2\frac{q}{m}n^{\sigma}\frac{dn^{\nu}}{ds}\left(\partial_{\sigma}F_{\,\,\,\,\nu}^{\mu}-F_{\,\,\,\,\beta}^{\mu}\Gamma^{\beta}_{\sigma\nu}\right).\label{n1}
\end{eqnarray}
As it clears, the left-hand side of the second-order geodesic deviation equation (\ref{n1}) is same to the left-hand side of equation (\ref{3}). As in the case of the second-order geodesic deviation, the higher-order geodesic deviation equations have the same left-hand side and different right-hand side. A non-manifest covariant form of the third-order geodesic deviation equation is given in appendix 1.

The successive approximations to the exact geodesic (b) have been shown in figure 1. Lines (c) and (d) represent the first-order approximation i.e. $x^{\mu}(s, p)=x^{\mu}(s, p_{0})+(p-p_{0})\frac{\partial x^{\mu}}{\partial p}|_{p_{0}}$, and the second-order approximation i.e. $x^{\mu}(s, p)=x^{\mu}(s, p_{0})+(p-p_{0})\frac{\partial x^{\mu}}{\partial p}|_{ p_{0}}+\frac{1}{2!}(p-p_{0})^{2}\frac{\partial ^{2} x^{\mu}}{\partial p^{2}}|_{p_{0}}$, respectively.

In the next section, we are going to obtain the components of $n^{\mu}$ from the first-order geodesic deviation, equation (\ref{3}), for a circular orbit of charged particles. Then by substituting them into equation (\ref{n1}), we can solve the second-order geodesic deviation equations, $b^{\mu}$. Finally, by substituting $n^{\mu}$ and $b^{\mu}$ into equation (\ref{6}), we will find the relativistic trajectory of charged particles in Reissner-Nordstrom space-time.

\section{The first-order geodesic deviation}
\subsection{Circular orbits in Reissner-Nordstrom metric}
The Reissner-Nordstrom metric is a static exact solution of the Einstein-Maxwell equations which describes the space-time around a spherically non-rotating charged source with mass $M$ and charge $Q$ (in the natural coordinate with $c=1$ and $G=1$)
\begin{equation}
ds^{2}=-B(r)dt^{2}+\frac{1}{B(r)}dr^{2}+r^{2}(d\theta^{2}+sin^{2}(\theta)d\varphi^{2}),\label{10}
\end{equation}
where
$$B(r)=1-\frac{2M}{r}+\frac{Q^2}{r^2}.$$
Also, the vector potential and the electromagnetic field of Maxwell's equations is \cite{Balakin}
\begin{equation}
A=A_{\mu}dx^{\mu}=-\frac{Q}{4\pi r}dt, \,\,\, F=dA=\frac{Q}{4\pi r^{2}}dr\wedge dt.\label{11}
\end{equation}
By assuming that $M^2>Q^2$, we are going to obtain the equation of motion for test particles which have mass $m$ and charge $q$. Now, we consider a circular orbit with a constant radius $R$. On the other hand, we know that the angular momentum of particles which are bounded to the spherically symmetric condition is limited to the equatorial plane. For this purpose and for simplicity, we limit the space to the plane of $\theta=\frac{\pi}{2}$ in which the angular momentum is in the $z$ direction. By using of the Euler-Lagrange equation, one can lead to the following constants of motion
\begin{equation}
\frac{d\varphi}{ds}=\frac{l}{r^{2}},\label{12}
\end{equation}
\begin{equation}
\frac{dt}{ds}=\frac{\varepsilon-\frac{qQ}{4\pi mr}}{1-\frac{2M}{r}+\frac{Q^{2}}{r^{2}}},\label{13}
\end{equation}
where $l=\frac{J}{m}$ is the angular momentum per unit mass,
$\dot{\varphi}=\omega$ is the angular velocity and $\varepsilon$ is the energy per unit mass.

Finally, from equations  (\ref{10}), (\ref{12}) and (\ref{13}) one obtains two constraints namely, the conservation of the absolute four-velocity and the radial acceleration. Now, due to the fact that the radius of the circular orbit is constant ($r=R$), two mentioned constraints vanish at all times and this creates two relations between $R$, $l$, and $\varepsilon$ as follows
\begin{equation}
\left(\varepsilon -\frac{qQ}{4\pi MR}\right)^{2} = \left(1-\frac{2M}{R}+\frac{Q^{2}}{R^{2}}\right)\left(1+\frac{l^2}{R^2}\right),\label{14}
\end{equation}
and
\begin{equation}\label{15}
\left[\frac{l^{2}}{R}-M\left(1+\frac{3l^{2}}{R^{2}}\right)+\frac{Q^{2}}{R}\left(1+\frac{2l^{2}}{R^{2}}\right)\right]^{2}=\left(\frac{qQ}{4\pi m}\right)^{2}\left(1+\frac{l^{2}}{R^{2}}\right)\left(1-\frac{2M}{R}+\frac{Q^{2}}{R^{2}}\right).
\end{equation}
As we expect that by eliminating charge, all obtained equations reduce to the similar equations in the Schwarzschild metric. In summary, we obtain the following four-velocity vector for a circular orbit with radius $R$ in an equatorial plane
\begin{equation}
u^{t}=\frac{dt}{ds}=\frac{\varepsilon -\frac{qQ}{4\pi m R}}{1-\frac{2M}{R}+\frac{Q^{2}}{R^{2}}}, \,\,\,\, u^{r}=\frac{dr}{ds}=0, \,\,\,\, u^{\theta}=\frac{d\theta}{ds}=0, \,\,\,\, u^{\varphi}=\frac{d\varphi}{ds}=\omega_{0}=\frac{l}{R^{2}}.\label{16}
\end{equation}
In the next subsection, we obtain the orbital motion by using the higher-order geodesic deviation method and compare the results with the perturbation
method.

\subsection{First-order geodesic deviation around the circular orbits}
Now let us calculate the first-order geodesic deviation for the components $n^t$, $n^r$, $n^{\theta}$ and $n^{\phi}$ , by using of equation (\ref{3}) in a matrix form
\begin{equation}
\left( \begin{array}{cccc} m_{11} & m_{12} & m_{13} & m_{14}  \\  m_{21} & m_{22} & m_{23} & m_{24}\\  m_{31} & m_{32} & m_{33} & m_{34}
\\  m_{41} & m_{42} & m_{43} & m_{44} \end{array} \right)\left( \begin{array}{cc}n^{t}\\ n^{r}\\n^{\theta} \\ n^{\varphi}\end{array} \right)
=\left( \begin{array}{cc}0\\0\\ 0 \\0\end{array}\right),\label{17}
\end{equation}
where the matrix elements are given by
$$
m_{11} = \frac{d^2}{ds^2},\hspace{1 cm}m_{12}=\frac{2R\varepsilon\left(\frac{M}{R}-\frac{Q^2}{R^2}\right)-\frac{qQ}{4\pi m}\left(1-\frac{Q^2}{R^2}\right)}{R^2\left(1-\frac{2M}{R}+\frac{Q^2}{R^2}\right)^2}\frac{d}{ds},\hspace{1 cm}m_{13}=m_{14}=0,
$$
$$
m_{22}=\frac{d^2}{ds^2}-\frac{l^2}{R^4}\left(1-\frac{Q^2}{R^2}\right)+\frac{\left(-\frac{2M}{R}+\frac{6M^2}{R^2}+\frac{3Q^2}{R^2}-\frac{12MQ^2}{R^3}
+\frac{5Q^4}{R^4}\right)\left(\varepsilon-\frac{qQ}{4\pi m R}\right)^2}{R^2\left(1-\frac{2M}{R}+\frac{Q^2}{R^2}\right)^2}-\frac{q}{m}F^{r}_{t,r} u^t,
$$
$$
m_{21}=\frac{2}{R}\left(\frac{M}{R}-\frac{Q^2}{R^2}\right)\left(\varepsilon-\frac{qQ}{4 \pi m R}\right){\frac{d}{ds}}-\frac{q}{m}{F^{r}_{t}}
{\frac{d}{ds}},\hspace{1 cm}
m_{23}=0, \hspace{1 cm}m_{24}=-\frac{2l}{R}\left(1-\frac{2M}{R}+\frac{Q^2}{R^2}\right)\frac{d}{ds},
$$
$$
m_{31}=m_{32}=m_{34}=0, \hspace{1 cm}m_{33}=\frac{d^2}{ds^2}+\frac{l^2}{R^4},
$$
$$
m_{41}=m_{43}=0, \hspace{1 cm}m_{42}=\frac{2l}{R^3}\frac{d}{ds},\hspace{1 cm} m_{44}=\frac{d^2}{ds^2}.
$$
As can be seen, the geodesic deviation equation of $\theta$ component represents a harmonic oscillator equation with the angular frequency of $\omega_{\theta}=\omega_0=\frac{l}{R^2}$. So we consider $n^{\theta}$ as follow
\begin{equation}
n^{\theta}(s) = n_{0}^{\theta} \cos(\omega_{0}s),\label{18}
\end{equation}
which is similar to the Schwarzschild case. So in this case we can neglect this solution ($n^{\theta}=0$), because the new plane of orbit is a new one inclined, or just a change of coordinate system \cite{Colistete}. Now, by eliminating the derivatives of $n^{t}$ and $n^{\phi}$ in the differential equation of $n^{r}$, we obtain the following oscillating equation
\begin{equation}
\frac{d^2 n^{r}}{ds^2}+w^2 n^{r}=0,\label{19}
\end{equation}
with the characteristic frequency
\begin{equation}
\omega^{2}=\omega_{0}^{2}\left(1-\frac{6M}{R}+\frac{Q^{2}}{MR}+\frac{3 Q^2}{R^2}+\cdots\right).\label{20}
\end{equation}
By considering $n_0^r>0$, we choose the following solution for $n^{r}$
\begin{equation}
n^{r} = -n^{r}_{0} \cos(\omega s).\label{21}
\end{equation}
Also, from the $n^{t}$ and $n^{\varphi}$ geodesic deviation equations, the solutions for $n^{t}$ and $n^{\varphi}$ are given by
\begin{equation}
n^{t} = n^{t}_{0} \sin(\omega s),\label{22}
\end{equation}
\begin{equation}
n^{\varphi} = n^{\varphi}_{0} \sin(\omega s),\label{23}
\end{equation}
where the amplitudes depend on $n_{0}^r$
\begin{equation}
n_{0}^{t}=\frac{2\sqrt{MR-{Q^{2}}}}{{R}(1-\frac{2M}{R}+\frac{Q^{2}}{R^{2}})\sqrt{1-\frac{6M}{R}+\frac{Q^{2}}{MR}}}n^{r}_{0},\label{24}
\end{equation}
\begin{equation}
n_{0}^{\varphi}=\frac{2}{R\sqrt{1-\frac{6M}{R}+\frac{Q^{2}}{MR}}}n^{r}_{0}.\label{25}
\end{equation}
In this way, the trajectory and the law of motion are obtained by
\begin{equation}
r=R-n^{r}_{0}\cos(\omega s),\label{26}
\end{equation}
\begin{equation}
\varphi = \omega_{0} s+n^{\varphi}_{0}\sin(\omega s),\label{27}
\end{equation}
\begin{equation}
t=\frac{\varepsilon - \frac{qQ}{4\pi m R}}{1-\frac{2M}{R}+\frac{Q^{2}}{R^{2}}}s+n^{t}_{0}\sin(\omega s),\label{28}
\end{equation}
where the argument phase of the cosine function is taken by $s=0$ for perihelion and $s=\frac{\pi}{\omega}$ for aphelion. Now, (\ref{26}) can be written as
\begin{equation}
r = R\left[1-\frac{n^{r}_{0}}{R}\cos(\omega s)\right].\label{29}
\end{equation}
By direct solution of the Euler-Lagrange equations, the trajectory of motion for particles is obtained in terms of centrifugal inertia \cite{Kepler}
\begin{equation}
r(t)=\frac{a(1-e^{2})}{1+e \cos(\omega_0 t)}\simeq a\left[1-e\cos(\omega_0 t)\right].\label{30}
\end{equation}
Obviously, equations (\ref{29}) and (\ref{30}) show that we have the same results. It means that if we bring up the eccentricity $e$ to $\frac{n^{r}_{0}}{R}$ and the semi-major axis $a$ to $R$, the same results are extracted, but there is also a difference that the
circular frequency, $\omega$, is lower than the circular frequency of the unperturbed circular motion, $\omega_0$. So, if the circular frequency decreases, the period increases. Then we obtain an expression for the periastron shift per one revolution as
\begin{equation}
\bigtriangleup\varphi = 2\pi \left(\frac{\omega_0}{\omega}-1\right) =  2\pi \left(\frac{3M}{R}+\frac{27}{2}\frac{M^{2}}{R^{2}}+\frac{135}{2}\frac{ M^3}{R^3}-\frac{Q^2}{2 M R}-\frac{6 Q^{2}}{R^{2}}+\cdots\right).\label{31}
\end{equation}
It can be seen from above equation that the charge parameter, $Q$, decreases the perihelion advance. In the perturbation method (Einstein's method), the orbital motion for charged particles moving in the equatorial plane of the Reissner-Nordstrom source is given by \cite{14}
\begin{equation}
\bigtriangleup\varphi = \frac{6\pi M}{R}-\frac{\pi Q^2}{MR},\label{311}
\end{equation}
comparing equation (\ref{31}) with equation (\ref{311}) shows that the presented method can be used in the vicinity of very massive and compact objects which is having a non-negligible ratio of $\frac{M}{R}$.

When the source is neutral and for the small values of $\frac{M}{R}$, equation (\ref{31}) reduces to the standard formula for Perihelion advance of planets \cite{Weinberg}. If we also compare equation (\ref{31}) to equation (\ref{1}), it is clear that in the first-order deviation, we hold only the terms up to $e^2$. In order to obtain the $\bigtriangleup\varphi$ for the higher values of the eccentricity, we must go beyond the first-order deviation equations. Therefore in the next section, we solve the second-order geodesic deviation equations in Reissner-Nordstrom space-time.

\section{The second-order geodesic deviation}
In this section, by using the first-order geodesic deviation equation and inserting equations (\ref{21}), (\ref{22}) and (\ref{23}) into equation (\ref{8}) and also doing a set of hard calculations, a linear equations system for the second-order geodesic deviation vector $b^{\mu}$ is obtained
\begin{equation}
\left( \begin{array}{cccc} m_{11} & m_{12} & m_{14}  \\  m_{21} & m_{22} & m_{24}
\\  m_{41} & m_{42} & m_{44} \end{array} \right)\left( \begin{array}{cc} b^{t}\\ b^{r}\\ b^{\varphi}\end{array} \right)=(n^{r}_{0})^{2}\left( \begin{array}{cc}C^{t}+C_{q}^{t}\\ C^{r}+C_{q}^{r} \\ C^{\varphi}+C_{q}^{\varphi}\end{array} \right),\label{322}
\end{equation}
where the constants $C^{t}$, $C^{t}_{q}$, $C^{r}$, $C^{r}_{q}$, and $C^{\varphi}$, $C^{\varphi}_{q}$ contain quantities depending on $M$, $R$, $\omega$ , $\omega_{0}$, $q$ and $Q$
\begin{equation}
C^{t}=-\frac{6M\sqrt{MR-Q^2}(2-\frac{7M}{R}+\frac{31Q^2}{3R^2}-\frac{5Q^2}{3MR}-\frac{4Q^4}{3MR^3}-\frac{Q^4}{M^2R^2})
}{R^5(1-\frac{2M}{R}+\frac{Q^2}{R^2})(1-\frac{3M}{R}+\frac{2Q^2}{R^2})\sqrt{1-\frac{6M}{R}+\frac{Q^2}{MR}}}\sin(2ws),\label{33}
\end{equation}
\begin{eqnarray}
C^{r}&=&-\frac{3M\left[6-\frac{27M}{R}+\frac{6M^2}{R^2}+\frac{158Q^2}{3R^2}-\frac{22Q^2}{3MR}-\frac{14MQ^2}{R^3}-\frac{16Q^4}{3R^4}-\frac{4Q^4}{M^2R^2}\right]
}{2R^4\left(1-\frac{3M}{R}+\frac{2Q^2}{R^2}\right)\left(1-\frac{6M}{R}+\frac{Q^2}{MR}\right)}\cos(2\omega s)\nonumber\\
&+&\frac{3M\left[2-\frac{5M}{R}+\frac{18M^2}{R^2}+\frac{6Q^2}{R^2}-\frac{10Q^2}{3MR}-\frac{34MQ^2}{R^3}
+\frac{4Q^4}{M^2R^2}\right]}{2R^4\left(1-\frac{3M}{R}+\frac{2Q^2}{R^2}\right)\left(1-\frac{6M}{R}+\frac{Q^2}{MR}\right)},
\end{eqnarray}
\begin{eqnarray}
C^{\varphi}=-\frac{6M\omega_{0}\left[1-\frac{3M}{R}+\frac{2M^2}{R^2}+\frac{5Q^2}{R^2}-\frac{4Q^2}{3MR}-
\frac{8MQ^2}{3R^3}-\frac{Q^4}{R^4}-\frac{Q^4}{MR^3}\right]}{\omega R^5\left(1-\frac{3M}{R}+\frac{2Q^2}{R^2}\right)\left(1-\frac{2M}{R}+\frac{Q^2}{R^2}\right)}\sin(2\omega s),
\end{eqnarray}
and
\begin{eqnarray}
C^{t}_{q}=\frac{qQ\sqrt{\frac{M}{R}-\frac{Q^2}{R^2}}\sqrt{1-\frac{6M}{R}+\frac{Q^2}{R^2}}\left[\frac{3M}{R}-\frac{31M^2}{2R^2}+\frac{15M^3}{R^3}
+\frac{Q^2}{R^2}+\frac{3MQ^2}{R^3}-\frac{7M^2Q^2}{R^4}\right]}{4\pi mMR^3\sqrt{1-\frac{3M}{R}+\frac{2Q^2}{R^2}}\left(1-\frac{2M}{R}+\frac{Q^2}{R^2}\right)^2\left(1-\frac{6M}{R}+\frac{Q^2}{MR}\right)}\sin(2\omega s),
\end{eqnarray}
\begin{eqnarray}
C^{r}_{q}&=&\frac{qQ\sqrt{1-\frac{3M}{R}+\frac{2Q^2}{R^2}}\left[\frac{7M}{R}-\frac{61M^2}{R^2}+\frac{169M^3}{R^3}-\frac{150M^4}{R^4}+\frac{3Q^2}{R^2}
+\frac{11MQ^2}{R^3}-\frac{130M^2Q^2}{R^4}+\frac{198M^3Q^2}{R^5}\right]}{4\pi mMR^3\left(1-\frac{3M}{R}+\frac{2Q^2}{R^2}\right)\left(1-\frac{6M}{R}+\frac{Q^2}{MR}\right)\left(1-\frac{2M}{R}+\frac{Q^2}{R^2}\right)}\cos(2\omega s)\nonumber\\
&-&\frac{qQ\sqrt{1-\frac{3M}{R}+\frac{2Q^2}{R^2}}\left[\frac{M}{R}+\frac{5M^2}{R^2}-\frac{45M^3}{R^3}+\frac{54M^4}{R^4}-\frac{3Q^2}{R^2}+\frac{21MQ^2}{R^3}-\frac{2M^2Q^2}{R^4}-\frac{54M^3Q^2}{R^5}\right]}{4\pi mMR^3\left(1-\frac{3M}{R}+\frac{2Q^2}{R^2}\right)\left(1-\frac{6M}{R}+\frac{Q^2}{MR}\right)\left(1-\frac{2M}{R}+\frac{Q^2}{R^2}\right)},
\end{eqnarray}
\begin{equation}
C^{\varphi}_{q}=0.
\end{equation}

Here we do not have used any approximation in $C^i$, $(i=r, \theta,\phi)$ but in what follows we neglect terms of higher order of the small parameters $\frac{M}{R}$, $\frac{Q}{M}$ and $\frac{q}{m}$. Solving the matrix equation (\ref{8}) for $b^{\mu}$ is similar to the approach used in the previous section (for the first-order geodesic deviation vector $n^{\mu}$) which contains the terms with characteristic frequency $\omega$. Here we are only interested in a particular solution because of the oscillating general solution with the angular frequency $\omega$ already take into account for $n^{\mu}(s)$. The particular solution of the above equation which is containing the oscillating terms with the angular frequency $2\omega$, the linear terms in the proper time $s$ and constants. To obtain the trajectory $x^{\mu}$ according to the equation (\ref{7}), we need to calculate $\frac{1}{2}\delta^{2}x^{\mu}$. Also for $x^{\mu}$, the perihelion is extracted by $\omega s=2k\pi$ and the aphelion is derived by $\omega s=(1+2k)\pi$ where $k\in Z$.

In appendix 2, we have put the particular solution of the above equation, $b^{\mu}$, the second-order geodesic deviation $\delta^{2}x^{\mu}$, and the semi-major axis $a$ and eccentricity $e$, respectively.

Finally, successive approximation brings us to trajectory by substituting $s(\varphi)$ to $\varphi(s)$
\begin{eqnarray}
\frac{r}{R}&=&1-\left(\frac{n_{0}^{r}}{R}\right) \cos\left(\frac{\omega}{\omega_{0}}\varphi\right)+\left(\frac{n_{0}^{r}}{R}\right)^{2}\biggr[
{\frac{\left(3-\frac{5M}{R}-\frac{30M^{2}}{R^{2}}+\frac{72M^{3}}{R^{3}}
+\frac{7Q^{2}}{R^{2}}-\frac{7Q^{2}}{MR}\right)}{2\left(1-\frac{2M}{R}+\frac{Q^{2}}{R^{2}}\right)\left(1-\frac{Q^{2}}{MR}\right)\left
(1-\frac{6M}{R}+\frac{Q^{2}}{MR}\right)^{2}}
}\nonumber\\
&+&\frac{\left(1-\frac{7M}{R}+\frac{10M^{2}}{R^{2}}+\frac{61Q^{2}}{2R^{2}}-\frac{8Q^{2}}{3MR}\right)}{2\left(1-\frac{2M}{R}
+\frac{Q^{2}}{R^{2}}\right)\left(1-\frac{6M}{R}+\frac{Q^{2}}{MR}\right)}
\cos\left(\frac{2\omega}{\omega_{0}}\varphi\right)
\nonumber\\
&+&\frac{\frac{3}{2}Qq}{32\pi M m\left(1-\frac{Q^{2}}{MR}\right)\left(1-\frac{2M}{R}\right)\left(1-\frac{3M}{R}\right)^{\frac{3}{2}}\left(1-\frac{6M}{R}+\frac{Q^2}{MR}\right)^{2}}
\nonumber\\
&+&\frac{\frac{19}{2}Qq}{32\pi Mm\left(1-\frac{Q^{2}}{MR}\right)\left(1-\frac{2M}{R}\right)\left(1-\frac{3M}{R}\right)^{\frac{3}{2}}\left(1-\frac{6M}{R}+\frac{Q^2}{MR}\right)^{2}} \cos\left(\frac{2\omega}{\omega_{0}}\varphi\right)\label{43}
\biggr]+\cdots
\end{eqnarray}
In the Schwarzschild limit, we have an elliptical orbit with \cite{Kerner}
\begin{eqnarray}
a&=&R+\frac{(n_0^{r})^2}{R}\biggr[\frac{(2-\frac{9M}{R}+\frac{11M^2}{R^2}+\frac{6M^3}{R^3})}{(1-\frac{2M}
{R})(1-\frac{6M}{R})^2}
\biggr],\label{44}
\end{eqnarray}
\begin{eqnarray}
e&=&\frac{n_{0}^{r}(1-\frac{2M}{R})
(1-\frac{6M}{R})^{2}}
{R(1-\frac{2M}{R})(1-\frac{6M}{R})^{2}+\frac{(n_0^{r})^2}{R}
(2-\frac{9M}{R}+\frac{11M^2}{R^2}+\frac{6M^3}{R^3})}=\frac{n_0^r}{R}+{\cal O}\left(\frac{(n_0^r)^3}{R^3}\right)
.\label{45}
\end{eqnarray}
Also, for the Schwarzschild case the shape of the orbit is described up to second-order of ($\frac{n_0^r}{R}$) as
\begin{eqnarray}
{\frac{r(\varphi)}{R}}&=& 1-\left(\frac{n_0^r}{R}\right)\cos\left(\frac{\omega}{\omega _0}\varphi\right)
+\left(\frac{n_0^r}{R}\right)^2\biggl[\frac{3-\frac{5M}{R}-\frac{30M^2}{R^2}+\frac{72M^3}{R^3}}{2(1-\frac{2M}{R})(1-\frac{6M}{R})^2}+\frac{(1-\frac{5M}{R})}{2(1-\frac{6M}{R})}\cos\left(\frac{2\omega}{\omega _0}\varphi\right)\biggr]+\cdots.
\end{eqnarray}
which is in agreement with equation (62) of reference \cite{Kerner}.

\section{Third-order geodesic deviation and Poincar\'{e}-Lindstedt's method}
In the previous section, we have calculated the trajectory of charged particles up to second-order. To find a more accurate trajectory, we need to obtain the higher-order terms of expansion (\ref{7}). Using the first and second-order
solutions and third-order equation (\ref{n3}) for $\delta^3x^{\mu}$, we have
\begin{eqnarray}
\left( \begin{array}{cccc} m_{11} & m_{12} & m_{14}  \\  m_{21} & m_{22} & m_{24}
\\  m_{41} & m_{42} & m_{44} \end{array} \right)\left( \begin{array}{cc} \delta^3{t}\\ \delta^3{r}\\\delta ^3{\varphi}\end{array} \right)=\epsilon^{3}\left( \begin{array}{cc}D^{t}\\ D^{r} \\ D^{\varphi}\end{array} \right),\label{t1}
\end{eqnarray}
where $m_{ij}$ are defined in equation (\ref{17}) and the coefficients $D^i_{0}$, $D^i_{0q}$, $D^i_{1}$, $D^i_{1q}$, $D^i_{3}$, $D^i_{3q}$, ($i=t, r, \varphi$) are functions of $M$, $R$, $q$, $Q$
$$
D^{t}=(D^{t}_{1}+D^{t}_{1q})\cos(\omega s)+(D^{t}_{3}+D^{t}_{3q})\cos(3\omega s)+D^{t}_{0}+D^{t}_{0q},
$$
$$
D^{r}=(D^{r}_{1}+D^{r}_{1q})\cos(\omega s)+(D^{r}_{3}+D^{r}_{3q})\cos(3\omega s)+D^{r}_{0}+D^{r}_{0q},
$$
$$
D^{\varphi}=(D^{\varphi}_{1}+D^{\varphi}_{1q})\cos(\omega s)+(D^{\varphi}_{3}+D^{\varphi}_{3q})\cos(3\omega s)+D^{\varphi}_{0}+D^{\varphi}_{0q}.
$$
As one can see the right-hand side of equations (\ref{t1}) have a frequency that is same as the eigenvalues of the differential matrix in the left-hand side (resonant terms). This makes a new problem i.e. infinite solution for $\delta^3r$ which is called the secular term (growing without bound). For avoiding these unbounded deviations we use the Poincar\'{e}'s method. In this method by replacing $\omega$ by infinite series in power of the infinitesimal parameter $\epsilon=\frac{n_0^r}{R}$ as
\begin{eqnarray}
\omega\rightarrow\omega_p=\omega+\epsilon\omega_1+\epsilon^2\omega_2+\epsilon_3\omega^3+\cdots,\label{t2}
\end{eqnarray}
the correction frequencies $\omega_1, \omega_2, \omega_3,\cdots$ can be chosen such that the Poincar\'{e}'s resonances vanish. By considering a differential equation for $x^{\mu}$ as
\begin{eqnarray}
&&\frac{d^2}{ds^2}\left(\delta r+\frac{1}{2}\delta^2r+\frac{1}{6}\delta^3r\right)+\omega^2\left(\delta r+\frac{1}{2}\delta^2r+\frac{1}{6}\delta^3r\right)=C^{r}_0+C^{r}_{0q}+(C^{r}+C^{r}_{q})\cos(2\omega_p s)\nonumber\\
&+&(D^{r}_{1}+D^{r}_{1q})\cos(\omega_p s)+(D^{r}_{3}+D^{r}_{3q})\cos(3\omega_p s)+D^{r}_{0}+D^{r}_{0q}.\label{t3}
\end{eqnarray}
Now, by developing both of the sides in terms of a series of the parameter $\epsilon$, for avoiding the secular terms, we find some algebraic relations on $\omega_1, \omega_2, \omega_3,\cdots$. In the Schwarzschild limit, we have \cite{Colistete}
\begin{eqnarray}
\omega_p=\frac{M^{1/2}\sqrt{1-\frac{6M}{R}}}{R^{3/2}\sqrt{1-\frac{3M}{R}}}-\epsilon^2
\frac{3M^{3/2}(6-\frac{37M}{R})}{4R^{5/2}\sqrt{1-\frac{3M}{R}}(1-\frac{6M}{R})^{3/2}},\label{t4}
\end{eqnarray}
where $\epsilon^2=\frac{(n_0^r)^2}{R^2}$. The resonant terms will also appear at the fifth-order approximation, by terms $\cos^5(ws),\sin^3(ws) \cos^2(ws)$, etc, this problem can be solved in a similar way.

Finally, we note that the electric charge of any celestial body being practically close to zero anyway. Therefore, it is worth to investigate the geodesic deviation and higher-order geodesic deviations in a more realistic background such as the Schwarzschild metric in a strong magnetic dipole field or magnetized black holes \cite{n1, n2, n3}. The study of them will be the subject of the future investigations.

\section{Conclusion and discussion}
Many of significant successes in general relativity are obtained by approximation methods. One of the most important approximation scheme in general relativity is the post-Newtonian approximation; an expansion with a small parameter which is the ratio of the velocity of matter to the speed of light. A novel approximation method was also proposed by Kerner et. al. which is based on the world-line deviations \cite{Kerner}.

The calculation of the perihelion advance by means of the higher-order geodesic deviation method for neutral particles in different gravitational fields such as Schwarzschild and Kerr metric was first studied in several papers \cite{Kerner, Colistete}. In the present paper by using of the higher-order geodesic deviation method for charged particles \cite{Heydari-Fard}, we applied this approximation method to charged particles in the Reissner-Nordstrom space-time.

We first started with an orbital motion which is close to a circular orbit with constant angular velocity which is considered as zeroth-approximation (unperturbed circular orbital motion) with the orbital frequency $\omega_0$. In the next step, we solved the first and second-order deviation equations which reduced to a system of the second-order linear differential equations with constant coefficients. The solutions are harmonic oscillators  with characteristic frequency. From equation (\ref{322}), the first and second-order corrections are oscillating terms with angular frequency $\omega$ and $2\omega$, respectively.

Finally, we have obtained the new trajectory by adding the higher-order geodesic deviations (non-linear effects) to the circular one, equation (\ref{43}). The advantage of this approach is to get the relativistic trajectories of planets without using Newtonian and post-Newtonian approximations for arbitrary values of quantity ${M}/{R}$.

\section*{Acknowledgements}
We wish to express our thanks to the anonymous referees for their comments and suggestions that helped us significantly to improve this
 manuscript.

\section*{Appendix 1}
For solving the third-order geodesic deviation equation, we should invoke to the Poincare's method. For this purpose, it is better to write the third-order geodesic deviation as $\delta^{3}x^{\mu}$. The third-order geodesic deviation equation $\delta^{3}x^{\mu}
$ is related to the third-order geodesic deviation vector $h^{\mu}$
\begin{eqnarray}
\delta^{3}x^{\mu}=\epsilon^3\left[h^{\mu}-3\Gamma^{\mu}_{\lambda \rho} n^{\lambda} b^{\nu}+(\partial_{\kappa}\Gamma^{\mu}_{\lambda \nu}-2\Gamma^{\mu}_{\lambda \sigma}\Gamma^{\sigma}_{\kappa \nu})n^{\kappa}n^{\lambda}n^{\nu}\right],\label{n2}
\end{eqnarray}
where $h^{\mu}=\frac{Db^{\mu}}{Dp}$. We derive the third-order geodesic deviation equation as
\begin{eqnarray}
\frac{d^{2}\delta^{3}x^{\mu}}{ds^{2}}&+&\left(2\Gamma^{\mu}_{\lambda \nu}u^{\lambda}-\frac{q}{m}F^{\mu}_{\,\,\,\,\nu}\right)\frac{d\delta^{3}x^{\nu}}{ds}+\left(\partial_{\sigma}\Gamma^{\mu}_{\lambda \rho} u^{\lambda}u^{\rho}-\frac{q}{m}u^{\nu}\partial_{\sigma}F^{\mu}_{\,\,\,\,\nu}\right)\delta^{3}x^{\sigma}=\nonumber\\
&-&6\Gamma^{\mu}_{\lambda\rho}\frac{d\delta x^{\lambda}}{ds}\frac{d\delta^{2} x^{\rho}}{ds}-3\delta x^{\sigma}(\partial_{\tau}\partial_{\sigma}\Gamma^{\mu}_{\lambda\rho})u^{\lambda}\left(\delta^{2} x^{\tau}u^{\rho}+2\delta x^{\tau}\frac{d\delta x^{\rho}}{ds}\right)\nonumber\\
&-&6 (\partial_{\sigma}\Gamma^{\mu}_{\lambda\rho})\left(\delta x^{\sigma}u^{\lambda}\frac{d\delta^{2} x^{\rho}}{ds}
+\delta x^{\sigma}\frac{d\delta x^{\lambda}}{ds}\frac{d\delta x^{\rho}}{ds}+\delta^{2} x^{\sigma}u^{\lambda}\frac{d\delta x^{\rho}}{ds}\right)-\delta x^{\sigma}\delta x^{\tau}\delta x^{\nu}(\partial_{\tau}\partial_{\sigma}\partial_{\nu}\Gamma^{\mu}_{\lambda\rho})u^{\lambda}u^{\rho}\nonumber\\
&+&\frac{q}{m}\frac{dn^{\nu}}{ds}\left[ (\partial_{\sigma}F^{\mu}_{\,\,\,\,\nu})\delta^{2}x^{\sigma}+(\partial_{\sigma}\partial_{\tau}F^{\mu}_{\,\,\,\,\nu})n^{\sigma}n^{\tau}\right] +\frac{q}{m}n^{\sigma}\left[ \frac{d\delta^{2}x^{\nu}}{ds}(\partial_{\sigma}F^{\mu}_{\,\,\,\,\nu})+
(\partial_{\sigma}\partial_{\tau}F^{\mu}_{\,\,\,\,\nu})\delta^{2}x^{\tau}u^{\nu}\right],\label{n3}
\end{eqnarray}
by substituting $\delta^{3}x^{\mu}$ in term of $h^{\mu}$ into above equation, we obtain equation (72) for case $q=0$ \cite{Kerner}.
\section*{Appendix 2}
The second-order geodesic deviation vector $b^{\mu}$ is
\begin{eqnarray}
b^{t}&=&\frac{(n^{r}_{0})^{2}M\left(\varepsilon -\frac{qQ}{4\pi m R}\right)}{R^{3}\left(1-\frac{6M}{R}+\frac{Q^{2}}{M R}\right)\left(1-\frac{2M}{R}\right)^{2}}\biggr
[-\frac{3(2-\frac{5M}{R}+\frac{18M^{2}}{R^{2}}
-\frac{10}{3}\frac{Q^{2}}{MR})}{(1-\frac{6M}{R}+\frac{Q^{2}}{MR})}s
\nonumber\\&+&\frac{(2-\frac{13M}{R}+\frac{79}{6}\frac{Q^{2}}{MR}) \sin(2\omega s)}{{\omega}}
-\frac{6qQws+19qQ \sin(2\omega s)}{32\pi m M\omega(1-\frac{6M}{R}+\frac{Q^{2}}{M R})(1-\frac{2M}{R})(1-\frac{3M}{R})^{\frac{3}{2}}}\biggr],\label{36}
\end{eqnarray}
\begin{eqnarray}
b^{r}&=&\frac{(n^{r}_{0})^{2}M}{2R^2\left(\frac{M}{R}-\frac{Q^{2}}{R^{2}}\right)\left(1-\frac{6M}{R}+\frac{Q^2}{M R}\right)}\biggr
[\frac{3(2-\frac{5M}{R}+\frac{18M^{2}}{R^{2}}
-\frac{10}{3}\frac{Q^{2}}{MR})}{(1-\frac{6M}{R}+\frac{Q^{2}}{MR})}
\nonumber\\&+&{(2+\frac{5M}{R}-\frac{28}{3}\frac{Q^{2}}{MR}) \cos(2\omega s)}
+\frac{3qQ+19qQ \cos(2\omega s)}{16\pi m M(1-\frac{6M}{R}+\frac{Q^{2}}{M R})(1-\frac{2M}{R})(1-\frac{3M}{R})^{\frac{3}{2}}}\biggr],\label{36}
\end{eqnarray}
\begin{eqnarray}
b^{\varphi}&=&\frac{(n^{r}_{0})^{2}\omega_{0}M}{ R^{3}(\frac{M}{R}-\frac{Q^2}{R^2})(1-\frac{6M}{R}+\frac{Q^{2}}{M R})}\biggr
[-\frac{3(2-\frac{5M}{R}+\frac{18M^{2}}{R^{2}}
-\frac{10}{3}\frac{Q^{2}}{MR})}{(1-\frac{6M}{R}+\frac{Q^{2}}{MR})}s
\nonumber\\&+&\frac{(1-\frac{8M}{R}) \sin(2\omega s)}{{2\omega}}
-\frac{6qQs+qQ(31-\frac{196M}{R}) \sin(2\omega s)}{32\pi m M(1-\frac{6M}{R}+\frac{Q^{2}}{M R})(1-\frac{2M}{R})(1-\frac{3M}{R})^{\frac{3}{2}}}\biggr].
\label{36}
\end{eqnarray}
As explained in section 2 the second-order geodesic deviation, $(\delta^{2}x = b^{\mu}-\Gamma^{\mu}_{\nu\lambda} n^{\nu}n^{\lambda})$, are given by
\begin{eqnarray}
\delta^{2}t&=&\frac{(n_{0}^{r})^{2}M\left(\varepsilon-\frac{qQ}{4\pi mR}\right)}{R^{3}}\biggr[-\frac{3\left(2-\frac{5M}{R}+\frac{18M^{2}}{R^{2}}-\frac{10Q^{2}}{3MR}\right)}{\left(1-\frac{2M}{R}\right)^{2}\left
(1-\frac{6M}{R}+\frac{Q^{2}}{MR}\right)}s
+\frac{\left(2-\frac{15M}{R}+\frac{14M^{2}}{R^{2}}-\frac{79Q^{2}}{6MR}\right)\sin(2\omega s)}{\omega\left(1-\frac{2M}{R}\right)^{3}\left(1-\frac{6M}{R}+\frac{Q^{2}}{MR}\right)}
\nonumber\\&-&\frac{6qQ\omega+19qQ\sin(2\omega s)}{32\pi mM\omega\left(1-\frac{2M}{R}\right)^{3}\left(1-\frac{3M}{R}\right)^{\frac{3}{2}}\left(1-\frac{6M}{R}+\frac{Q^{2}}{MR}\right)^{2}}\biggr],\label{39}
\end{eqnarray}

\begin{eqnarray}
\delta^{2}r&=&\frac{(n_{0}^{r})^{2}M}{R^2\left(\frac{M}{R}-Q^{2}\right)\left(1-\frac{6M}{R}+\frac{Q^{2}}{MR}\right)}\biggr[\frac{\left(5-\frac{33M}{R}
+\frac{90M^{2}}{R^{2}}-\frac{72M^{3}}{R^{3}}+\frac{5Q^{2}}{R^{2}}-\frac{5Q^{2}}{MR}\right)}{\left(1-\frac{2M}{R}+\frac{Q^{2}}{R^2}\right)\left
(1-\frac{6M}{R}+\frac{Q^{2}}{MR}\right)}
\nonumber\\&+&\frac{\left(-1+\frac{9M}{R}+\frac{33Q^2}{2R^2}-\frac{19M^{2}}{R^2}-\frac{23}{2}\frac{MQ^{2}}{R^3}-\frac{8Q^{2}}{3MR}\right) \cos(2\omega s)}{\left(1-\frac{2M}{R}+\frac{Q^{2}}{R^2}\right)}
\nonumber\\&+&\frac{3qQ+ 19Qq\cos(2\omega s)}{32\pi mM\left(1-\frac{2M}{R}\right)\left(1-\frac{3M}{R}\right)^{\frac{3}{2}}\left(1-\frac{6M}{R}+\frac{Q^{2}}{MR}\right)}\biggr],\label{40}
\end{eqnarray}

\begin{eqnarray}
\delta^{2}\varphi &=&\frac{(n_{0}^{r})^{2}M\omega_{0}}{R^3\left(\frac{M}{R}-\frac{Q^{2}}{R^2}\right)\left(1-\frac{6M}{R}+\frac{Q^{2}}{MR}\right)}
\biggr[-\frac{3\left(2-\frac{5M}{R}+\frac{18M^{2}}{R^{2}}-\frac{10Q^{2}}{3MR}\right)}
{\left(1-\frac{6M}{R}+\frac{Q^{2}}{MR}\right)}s
\nonumber\\&+&\frac{\left(5-\frac{32M}{R}\right)\sin(2\omega s)}{2\omega}
-\frac{6qQs+qQ\left(31-\frac{196M}{R}\right)\sin(2\omega s)}{32\pi mM\left(1-\frac{2M}{R}\right)\left(1-\frac{3M}{R}\right)^{\frac{3}{2}}\left(1-\frac{6M}{R}+\frac{Q^{2}}{MR}\right)}\biggr],\label{41}
\end{eqnarray}
also, the semi-major axis $a$ and eccentricity $e$ are
\begin{eqnarray}
a=R+\frac{(n_0^{r})^2}{R}\biggr[\frac{(2-\frac{9M}{R}+\frac{11M^2}{R^2}+\frac{6M^3}{R^3}+\frac{15Q^2}{R^2}-\frac{13Q^2}{3MR}-\frac{235MQ^2}{4R^3})}{(1-\frac{2M}
{R}+\frac{Q^{2}}{R^2})(1-\frac{Q^2}{MR})(1-\frac{6M}{R}+\frac{Q^{2}}{MR})^2}\biggr],
\end{eqnarray}
\begin{eqnarray}
e&=&\frac{n_{0}^{r}(1-\frac{2M}{R}+\frac{Q^2}{R^2})
(1-\frac{Q^2}{MR})(1-\frac{6M}{R}+\frac{Q^2}{MR})^{2}}
{R(1-\frac{2M}{R}+\frac{Q^2}{R^2})(1-\frac{Q^2}{MR})(1-\frac{6M}{R}+\frac{Q^2}{MR})^{2}+\frac{(n_0^{r})^2}{R}
(2-\frac{9M}{R}+\frac{11M^2}{R^2}+\frac{6M^3}{R^3}+\frac{15Q^2}{R^2}-\frac{13Q^2}{3MR}-\frac{235MQ^2}{4R^3})}
,\label{42}
\end{eqnarray}
in which for massive central objects we have neglected all terms of order $\frac{qQ}{4\pi m M}$.

\end{document}